\documentclass{ws-procs9x6}

\input{epsf}

\def\CQG{{\it Class. Quantum Gravity} }

\def\PR{{\it Phys. Rev.} }
\def\PRL{{\it Phys. Rev. Lett.} }

\def\al{\alpha}
\def\be{\beta}
\def\ga{\gamma}
\def\de{\delta}

\def\th{\theta}

\def\ka{\kappa}
\def\la{\lambda}

\def\De{\Delta}

\def\Om{\Omega}

 \def\frac#1#2{{\textstyle{{#1}\over
{#2}}}} 
\def\lsim{\mathrel{\rlap{\lower4pt\hbox{\hskip1pt$\sim$}}
    \raise1pt\hbox{$<$}}} \def\gsim{\mathrel{\rlap{\lower4pt\hbox{\hskip1pt$\sim$}}
    \raise1pt\hbox{$>$}}}
\def\sqr#1#2{{\vcenter{\vbox{\hrule height.#2pt
         \hbox{\vrule width.#2pt height#1pt \kern#1pt
         \vrule width.#2pt}
         \hrule height.#2pt}}}} 

\def\beq{\begin{equation}}
\def\eeq{\end{equation}}
\def\beqa{\begin{eqnarray}}
\def\eeqa{\end{eqnarray}}

\begin{document}

\title{Scalar field models: from the Pioneer anomaly to astrophysical constraints}

\author{J. P\'aramos \footnote{\uppercase{W}ork partially
supported by the \uppercase{F}unda\c{c}\~{a}o para a
\uppercase{C}i\^{e}ncia e \uppercase{T}ecnologia, under the grant
\uppercase{BD}~6207/2001.}}

\address{Instituto Superior T\'ecnico, Departamento de
F\'{\i}sica, \\ Avenida Rovisco Pais 1, 1049-001 Lisboa,
Portugal\\E-mail: x\_jorge@netcabo.pt}

\maketitle

\abstracts{In this work we study how scalar fields may affect
solar observables, and use the constraint on the Sun's central
temperature to extract bounds on the parameters of relevant
models. Also, a scalar field driven by a suitable potential is
shown to produce an anomalous acceleration similar to the one
found in the Pioneer anomaly.}

\section{Introduction}

Scalar fields play a fundamental role in contemporary physics,
with applications ranging from particle physics to cosmology and
condensed matter. In cosmology, these include
inflation\cite{Olive}, vacuum energy evolving and quintessence
models, the Chaplygin gas dark energy-dark matter unification
model and dark matter candidates (for a complete list, see $[2]$
and references within). A scalar field has recently been suggested
as a possible explanation for the anomalous acceleration measured
by the Pioneer spacecraft\cite{Bertolami}. On an astrophysical
context, a scalar field could be present as, for example, a
mediating boson of an hypothetical fifth force. This can be
modelled by a Yukawa potential, given by $V_Y(r) = A e^{-mr} / r
$, where $A$ is the coupling strength and $m$ is the mass of the
field, setting the range of the interaction $\la_Y \equiv m^{-1}$.
The bounds on parameters $A$ and $\la_Y \equiv m^{-1}$  can be
found in Ref.\cite{Fischbach} and references therein. In this
work, a scalar field with an appropriate potential is shown to
account for the anomalous Pioneer acceleration. Also, the effects
of this and other scalar field models on stellar equilibrium is
reported.

\section{Scalar field induced Pioneer anomaly}

The hypothesis that the  Pioneer anomaly is due to new physics has
been steadily gaining momentum. Fundamentally, this arises because
is seems unfeasible that an engineering explanation, affecting
spacecrafts with different trajectories and designs, could lead to
similar effects. The lack of a simple explanation could hint that
this anomalous acceleration is the manifestation of a new force.

Given the broad range of applicability of theories with scalar
fields, it is possible that such an entity is responsible for the
anomaly\cite{Bertolami}. A natural candidate would be the radion,
a scalar field present due to oscillations of inter-brane distance
in braneworld models. However, this does not account for the
phenomena\cite{Bertolami}. Fortunately, one can put down a model
with a scalar field that doesn't stray away much from the already
accepted theories arising from cosmology. Specifically, we assume
a scalar field with dynamics ruled by a potential of the form $
V(\phi) = - A^2 \phi^{-2} $, where $A$ is a constant. The scalar
field obeys the equation of motion

\beq \phi''(r) + {2 \over r} \phi'(r) = 2 A^2 \phi^{-3} ~~,
\label{eqmotion} \eeq

\noindent and admits the solution

\beq \phi(r) = \left( {8 \over 3} \right)^{1 \over 4} \sqrt{ A r }
\equiv \be^{-1} r^{1 \over 2}~~. \label{solution} \eeq

The $0-0$ linearized Einstein's equations reads

\beq f''(r) + {2 \over r} f'(r) =  2\ka \left( 1 - {2 C \over r}
\right) A^2  \be^2 r^{-1} \label{00} \eeq

\noindent whose solution is

\beq f(r) = \sqrt{3 \over 2} {A \ka r \over 2} - \sqrt{6} AC \ka
~log \left({r \over 1~m} \right) ~~.\eeq

\noindent The resulting acceleration felt by a test body is given
by

\beq a_r = -{C \over r^2} - \sqrt{3 \over 2} {A \ka \over 4} +
\sqrt{3 \over 2} {A C \ka \over r} ~~. \label{alfa2} \eeq

The first term is the Newtonian contribution, and identifying the
second with the anomalous acceleration $a_A = 8.5 \times 10^{-10}~
m s^{-2} $, sets $ A = 3.26 ~ a_A / \ka = 4.7 \times
10^{42}~m^{-3}$. The last term is much smaller than the anomalous
acceleration for $ 4C/r \ll 1$, that is, for $r \gg 6~km$; it is
also much smaller than the Newtonian acceleration for $r \ll ~ 2.9
\times 10^{22}~km \approx 100~Mpc$. It has been
shown\cite{Bertolami} that the effect of this model on the Doppler
effect used to calculate the acceleration of the spacecrafts is
negligible, indicating that the anomaly is real, an not due to a
misinterpreted light propagation.

\section{Variable Mass Particle models}

The variable mass particle (VAMP) proposal\cite{Carroll} assumes
the presence of yet unknown fermions coupled to a dark-energy
scalar field with dynamics ruled by a monotonically decreasing
potential. Although this has no minima, the coupling of the scalar
field $\phi$ to this exotic fermionic matter yields an effective
potential of the form $V_{eff} (\phi) = V(\phi) + \la n_\psi
\phi$, where $n_\phi$ is the number density of fermionic VAMPs and
$\la$ is a Yukawa coupling. In the present work we approach a
potential of the quintessence-type form, $V(\phi) = u_0
\phi^{-p}$. As a result, the effective potential acquires a
minimum and the related vacuum expectation value (\textit{vev}) is
responsible for the mass term of the exotic VAMP. Since the number
density $n_\psi$ depends on the scale factor $a(t)$, this mass
varies on a cosmological timescale. A noticeable shortcoming of
VAMP models is the presence of weakly or unconstrained parameters:
the relative density $\Om_{\psi0}$, the scalar field coupling
constant and the potential strength.

One can attempt to overcome this drawback by assuming that all
fermions couple to the quintessence scalar field. This coupling
should not substitute the usual Higgs coupling, but merely add a
small correction to the Higgs-mechanism induced mass. Since the
effects of the quintessence scalar field coupling crucially depend
on the particle number density $n_\psi$, one expects this variable
mass term to play a more relevant role in a stellar environment
than in the vacuum. This hints that VAMP model parameters could be
constrained from stellar physics observables.

\subsection{The polytropic gas stellar model}

The polytropic gas model assumes an equation of state of the form
$P = K \rho^{n+1/n}$, where $n$ is the polytropic index, defining
intermediate processes between the isothermic and adiabatic
thermodynamical cases, and $K$ is the polytropic constant, which
depends on the star's mass $M$ and radius $R$. This model leads to
scaling laws for thermodynamical quantities, given by $\rho =
\rho_c \th^n(\xi)$, $ T = T_c \th(\xi)$ and $ P = P_c
\th(\xi)^{n+1}$, where $\rho_c$, $T_c$ and $P_c$ is the density,
temperature and pressure at the center of the star\cite{Bhatia}.
The dimensionless function $\th(\xi)$ depends on the dimensionless
variable $\xi$, related to the distance to the star's center by $r
= \al \xi$, where $\al$ also depends on the star's mass $M$ and
radius $R$. The hydrostatic equilibrium condition enables a
differential equation ruling the behavior of the scaling function
$\th(\xi)$, the Lane-Emden equation:

\beq {1 \over \xi^2} {\partial \over \partial \xi} \left(\xi^2
{\partial \th \over \partial \xi} \right) = -\th^n~~. \eeq

\noindent Since the physical characteristics of a star appear only
in the definitions of the constants $K$ and $\al$, its stability
is independent of these quantities, and different polytropic
indexes $n$ label different types of stars. This
scale-independence can be related to the homology symmetry
enclosed in the Lane-Emden equation. The boundary conditions of
this differential equation are, from the definition $\rho = \rho_c
\th(\xi)$, $\th(0)=1$; also, the hydrostatic equilibrium condition
implies that $\left|{d \th / d \xi} \right|_{\xi=0} = 0$. The
first solar model ever considered corresponds to a polytropic star
with $n=3$ and was studied by Eddington in 1926. Although somewhat
incomplete, this simplified model gives rise to relevant
constraints on the physical quantities.

The following results are based on the luminosity constraint on
the Sun's central temperature, $\De T_c/T_c \leq 0.4\%$. The
central temperature can be computed from $ T_c = Y_n \mu GM/R$,
where $k$ is the Boltzmann constant and $Y_n$ depends on the
star's mass $M$ and radius $R$, as well as on the dimensionless
quantity $\xi_1$, which is defined through $\th(\xi_1)=0$ and
signals the surface of the star.

\subsection{Results}

Assuming isotropy, a variable mass term leads to both a radial,
anomalous acceleration $a_A = \phi' / \phi < 0$ plus a
time-dependent drag force $ a_D = -\dot{\phi} / \phi <
0$\cite{astro}. The time-dependent component should vary on
cosmological timescales, and can thus can be absorbed in the usual
Higgs mass term. Hence, one considers only the perturbation to the
Lane-Emden equation given by the radial force. We start by
defining the dimensionless scalar field $\Phi \equiv \phi /
\phi_c^*$, with $ \phi_c^* = \rho_{crit} / 2 n_V$, where $\Om_{V}$
is the energy density due to the potential driving $\phi$ and
$\rho_{crit} \simeq 9.48 \times 10^{-28}~g~cm^{-3}$ is the
critical density. Assuming for simplicity a potential with $p=1$,
we obtain a perturbed Lane-Emden equation

\beq {1 \over \xi^2} {d \over d \xi} \left[ \xi^2 {d \th \over d
\xi}\right] = - \th^n(\xi) - {C_n \over U} {1 \over \Phi(\xi)}
\left[\Phi''(\xi) + {2 \over \xi} \Phi'(\xi) - {\Phi'^2(\xi) \over
\Phi(\xi) } \right]~~. \label{Lane-Emden-Phi} \eeq

\noindent where one has defined the dimensionless quantities

\beq U  \equiv {G M \over R c^2} = 2.12 \times 10^{-6} ~~,~~
C_n^{-1} \equiv (n+1) N_n^{n/(n+1)} W_n^{1 / (n+1)}~~, \eeq

\noindent with $ W_n = [4 \pi (n+1)\left({d \th \over d \xi}
\right)^2_{\xi_1}]^{-1}$.

Furthermore, one assumes that the scalar field is only weakly
perturbed in relation to the cosmological \textit{vev} of the
effective potential. Denoting this  small ``astrophysical''
contribution as $\Phi_a(\xi)$, one has $\Phi(\xi) = \Om_V / \la +
\Phi_a(\xi)$. Hence, the Klein-Gordon equation becomes

\beq  \Phi''_a(\xi) + {2 \over \xi} \Phi'_a(\xi) \approx {2 \al^2
\la n_V \over \rho_{crit}} {\rho_c \over \mu } \th^3(\xi) -{{2
\al^2 \la^2 n_V^2 \over \rho_{crit}}} \left[1-{ 2 \la \Phi_a(\xi)
\over \Om_V} \right] ~~, \eeq

\noindent inside the star, and

\beq \Phi''_a(\xi) + {2 \over \xi} \Phi'_a(\xi) \approx {2 \al^2
\la n_V^2 \over \rho_{crit}} -{{2 \al^2 \la^2 n_V^2 \over
\rho_{crit}}} \left[1-{ 2 \la \Phi_a(\xi) \over \Om_V} \right] ~~,
\eeq

\noindent in the outer region. In the above, $n_\psi \equiv n_V =
3 ~m^{-3} $ is the number density of fermions in the vacuum. $\mu$
is the mean molecular weight of Hydrogen. The boundary conditions
for the perturbed Lane-Emden are the same as in the unperturbed
case. For the scalar field, we assume both $\Phi(\xi)$ and its
derivative vanish beyond the Solar System (about $10^5~ AU$).
Following\cite{astro}, one gets that the maximum deviation $\De
T_c / T_c = 2.82 \times 10^{-8}$ occurs for $\Om_V=0.7$, $\la=2.82
\times 10^{-14}$. Hence, the luminosity constraint is always
respected as long as one assumes the bound arising from the
assumption that the ``astrophysical'' component of the scalar
field is much smaller than its cosmological \textit{vev}, $\la <
10^{-14}$.

\section{Yukawa potential induced perturbation}

One now looks at the hydrostatic equilibrium equation with a
Yukawa potential which, after a small algebraic manipulation,
implies the perturbed Lane-Emden equation

\beq {1 \over \xi^2} {d \over d \xi} {d \th \over d \xi}  = -\th^n
\left[1 + Ae^{-\ga \xi / \xi_1}\right] - \ga A C_n {d \th \over d
\xi} e^{- \ga \xi / \xi_1} ~~, \label{LEY} \eeq

\noindent where we have defined the dimensionless quantities

\beq C_n \equiv \left({n+1 \over 4 \pi}\right)^{1/2} N_n^{n/(n+1)}
W_n^{(1-n)/2(n+1)}~~,~~\ga \equiv m R \eeq

\noindent and $\xi_1$ signals the surface of the star (more
accurately, a surface of zero temperature, but the difference is
negligible. It can be shown that the boundary conditions are
unaffected by the perturbation\cite{astro}.

\begin{figure}

\epsfysize=10cm \epsffile{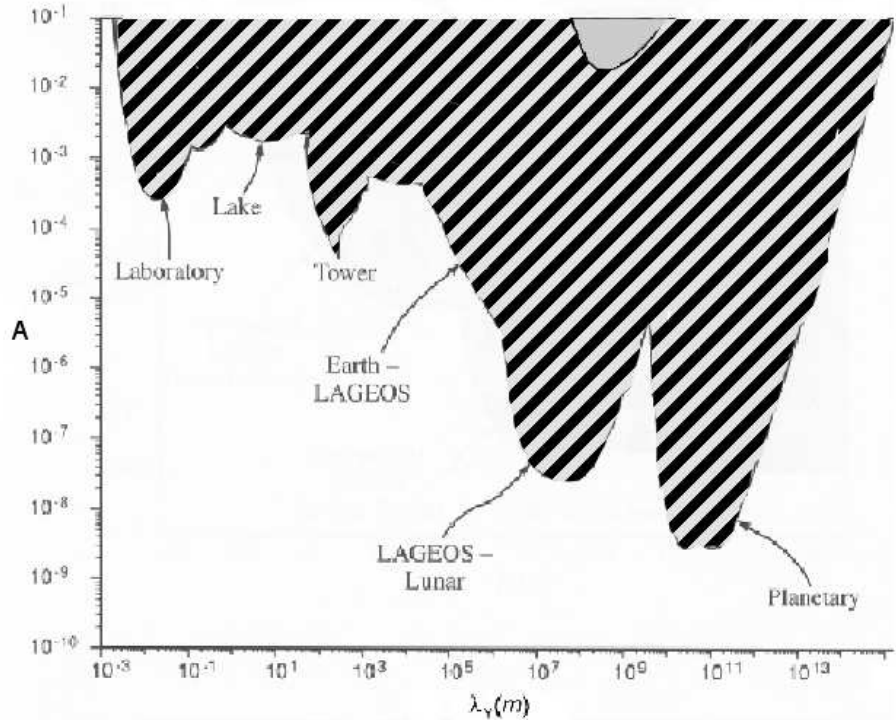} \caption{Exclusion plot
for the relative deviation from unperturbed central temperature
$T_c$, for $A$ ranging from $10^{-3}$ to $10^{-1}$, and $\ga$ from
$10^{-1}$ to $10$ (tip at the top), superimposed on the available
bounds.} \label{graph_exc_y}

\end{figure}

One can study the variation of the central temperature as a
function of $\la \equiv m^{-1}$ and $A$, and constraint these so
that $\De T_c / T_c \leq 0.4 \% $. The parameters were chosen for
Yukawa interactions in the range $0.1R < \la_Y < 10R$; the Yukawa
coupling $A$ was chosen so that the variation of $T_c$ is of the
same order $O(10^{-4})$) as the luminosity constraint. Numerical
integration of Eq. \ref{LEY} is then used to derive the exclusion
plot of Figure \ref{graph_exc_y}, superimposed on the accepted
bound\cite{Fischbach}. Notice that the central temperature is not
precisely known and it is clear that constraining its uncertainty
below $10^{-4}$ would yield a larger exclusion region in the
parameter space.

\subsection{The Pioneer anomaly}

Following a similar procedure to the one depicted in the above two
sections, one can prove that a constant, anomalous acceleration
$a_A$ inside the Sun yields a relative deviation of the central
temperature which scales linearly with $a_A$ as $\de T_c \sim
a_A/a_\odot$, where $a_\odot = 274~m.s^{-2}$\cite{astro}. Thus,
the bound $\de T_c <4 \times 10^{-3}$ is satisfied for values of
this constant anomalous acceleration up to $a_{Max} \sim 10^{-4}
a_\odot$. The reported value is then well within the allowed
region, and has a negligible impact on the astrophysics of the
Sun.

\section{Conclusion}

In this work we have developed a study of the impact of some
exotic physics models on stellar equilibrium, enabling the
extraction of constraints on the relevant parameters of the
theories. Also, a model exhibiting a scalar field with a
attractive quintessence-like potential is discussed, which can
account for the anomalous acceleration felt by the Pioneer 10/11,
Ulysses and Galileo spacecrafts.

\end{document}